\documentclass[aps,twocolumn,prd,dvipsnames,usenames,floatfix,superscriptaddress,nofootinbib,longbibliography,preprintnumbers]{revtex4-1}
\pdfoutput=1

\usepackage{amsmath,amssymb,bm}

\usepackage{fancyhdr} 
\usepackage{footnote}
\makesavenoteenv{tabular}
\makesavenoteenv{table}
\usepackage[utf8x]{inputenc}
\usepackage{units}
\usepackage[english]{babel}
\usepackage{xspace}
\usepackage{paralist}
\usepackage{amsmath} 
\usepackage{float}
\usepackage{amssymb} 
\usepackage{graphicx}
\usepackage{bm}
\usepackage{tabularx}
\usepackage{slashed}
\usepackage[final,textsize=footnotesize,color=ForestGreen!05]{todonotes}

\usepackage{subfig}
\usepackage[colorlinks=true,linktocpage=true,urlcolor=blue,citecolor=blue,linkcolor=blue]{hyperref}

\usepackage{dsfont}

\usepackage[justification=centerlast, format=plain, labelfont=bf,font=small]{caption}

\DeclareCaptionJustification{justified}{\leftskip=0pt \rightskip=0pt \parfillskip=0pt plus 1fil}

\renewcommand{\thefootnote}{\arabic{footnote}}

\makeatletter
\def\doauthor#1#2#3{%
  \ignorespaces#1\unskip
  \begingroup
   #3%
  \@if@empty{#2}{\@listcomma\endgroup{}{}}{\endgroup{\comma@space}{}\frontmatter@footnote{#2}}%
  \space \@listand
}%
\makeatother

\makeatletter
\def\@ssect@ltx#1#2#3#4#5#6[#7]#8{%
  \def\H@svsec{\phantomsection}%
  \@tempskipa #5\relax
  \@ifdim{\@tempskipa>\z@}{%
    \begingroup
      \interlinepenalty \@M
      #6{%
       \@ifundefined{@hangfroms@#1}{\@hang@froms}{\csname @hangfroms@#1\endcsname}%
       {\hskip#3\relax\H@svsec}{#8}%
      }%
      \@@par
    \endgroup
    \@ifundefined{#1smark}{\@gobble}{\csname #1smark\endcsname}{#7}%
  }{%
    \def\@svsechd{%
      #6{%
       \@ifundefined{@runin@tos@#1}{\@runin@tos}{\csname @runin@tos@#1\endcsname}%
       {\hskip#3\relax\H@svsec}{#8}%
      }%
      \@ifundefined{#1smark}{\@gobble}{\csname #1smark\endcsname}{#7}%
      \addcontentsline{toc}{#1}{\protect\numberline{}#8}%
    }%
  }%
  \@xsect{#5}%
}%
\makeatother

\makeatletter
\g@addto@macro\bfseries{\boldmath}
\makeatother
\graphicspath{{./}}

\begin{document}

\title{Blast from the past: constraints from the CHARM experiment on Heavy Neutral Leptons with tau mixing}
 \author{Iryna Boiarska}
 \affiliation{Discovery Center, Niels Bohr Institute, Copenhagen University, Blegdamsvej 17, DK-2100 Copenhagen, Denmark}
 \author{Alexey Boyarsky}
 \affiliation{Instituut-Lorentz for Theoretical Physics, Universiteit Leiden, Niels Bohrweg 2, 2333 CA Leiden, Netherlands}
 \author{Oleksii Mikulenko}
 \affiliation{Instituut-Lorentz for Theoretical Physics, Universiteit Leiden, Niels Bohrweg 2, 2333 CA Leiden, Netherlands}
 \author{Maksym Ovchynnikov}
\affiliation{Instituut-Lorentz for Theoretical Physics, Universiteit Leiden, Niels Bohrweg 2, 2333 CA Leiden, Netherlands}

\def\thefootnote{\hspace{0.7pt}}\footnotetext{\href{mailto:boiarska@nbi.ku.dk}{boiarska@nbi.ku.dk}}
\def\thefootnote{\hspace{0.4pt}}\footnotetext{\href{mailto:boyarsky@lorentz.leidenuniv.nl}{boyarsky@lorentz.leidenuniv.nl}}
\def\thefootnote{\hspace{0.7pt}}\footnotetext{\href{mailto:mikulenko@lorentz.leidenuniv.nl}{mikulenko@lorentz.leidenuniv.nl}}
\def\thefootnote{\hspace{0.pt}}\footnotetext{\href{mailto:ovchynnikov@lorentz.leidenuniv.nl}{ovchynnikov@lorentz.leidenuniv.nl}}
\setcounter{footnote}{0}
\def\thefootnote{\arabic{footnote}}

\begin{abstract}
\noindent We re-analyze the results of the searches for Heavy Neutral Leptons (HNLs) by the CHARM experiment. We study HNL decay channel $N\to e^{+}e^{-}\nu/\mu^{+}\mu^{-}\nu$ and show that, in addition to the constraints on the HNL's mixings with $\nu_e$ or $\nu_{\mu}$, the same data also implies limits on the HNLs that mix only with $\nu_{\tau}$ and have masses in the range $\unit[290]{MeV}<m_{N}\lesssim \unit[1.6]{GeV}$  - the region in the parameter space that was considered in the literature as a target for HNLs searches.
\end{abstract}
\maketitle

\section{Introduction and summary}
\label{sec:introduction}
The Standard Model (SM) of particle physics, despite its unprecedented success, has proven to be incomplete, encouraging searches for new  particles. Heavy Neutral Leptons, or HNLs, is a well-motivated extension of the SM that is capable of simultaneously explaining several long-standing problems: neutrino oscillations~\cite{Minkowski:1977sc}, dark matter~\cite{Asaka:2005an, Boyarsky:2018tvu, Boyarsky:2009ix}, and baryon asymmetry of the Universe~\cite{Asaka:2005pn, Asaka:2010kk, Asaka:2011wq, Shaposhnikov:2009zzb}. From the phenomenological point of view, such particles participate in weak interactions and behave as heavy neutrinos with interaction strength suppressed by the mixing angles $U_\alpha$, as compared to ordinary neutrinos $\nu_{\alpha}$ (see, e.g.~\cite{Bondarenko:2018ptm}). HNLs with the masses in the MeV-GeV range are searched for at the accelerators (see, e.g.~\cite{Beacham:2019nyx, Agrawal:2021dbo}), and may be constrained from cosmological observations as well~\cite{Boyarsky:2020dzc,Boyarsky:2021yoh,Sabti:2020yrt}. 

In the minimal models where two or three HNLs 
explain neutrino flavor oscillation data via a seesaw mechanism, each of the HNLs should have comparable mixings with all active neutrino flavors (see, e.g.~\cite{Bondarenko:2021cpc,Chrzaszcz:2019inj,Drewes:2018gkc} and references therein for discussion). Nevertheless, to obtain the accelerator bounds, it is convenient to consider HNLs that mix with only one flavor of active neutrinos, namely $\nu_\alpha$ (below, we denote such HNLs as $N_{\alpha}$). Limits on $U^2_\alpha$ derived for such simplified ``pure mixing model'' are conservative,
as the presence of additional mixing angles $U^2_{\beta\neq\alpha}\neq 0$ for the same particle would only increase the number of expected events in a given experiment at the lower bound of sensitivity.

\begin{figure*}[h!]
    \centering
    \includegraphics[width=0.6\textwidth]{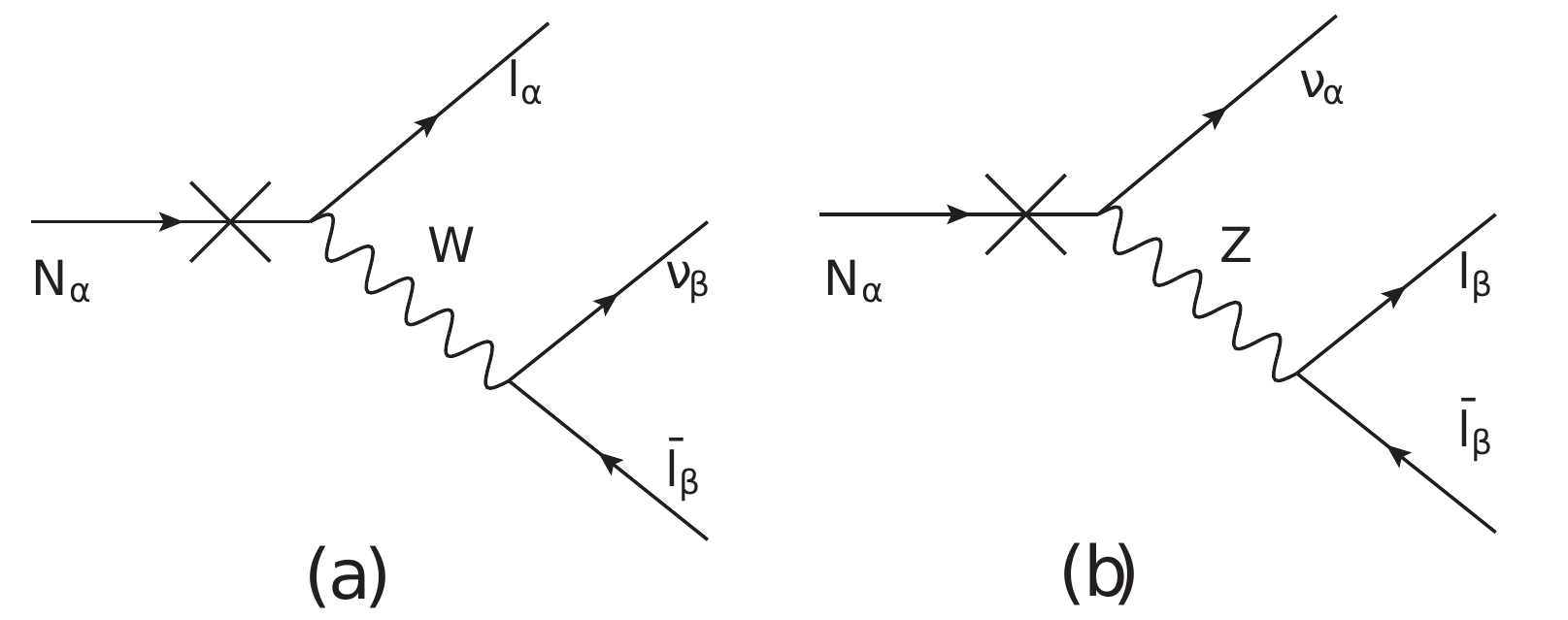}
    \caption{Diagrams of leptonic decays of an HNL that mixes purely with $\nu_{\alpha}$ via the charged (the left diagram) and the neutral current (the right diagram).}
    \label{fig:hnl-decays}
\end{figure*}

\begin{figure*}[h!]
    \includegraphics[width=0.68\textwidth]{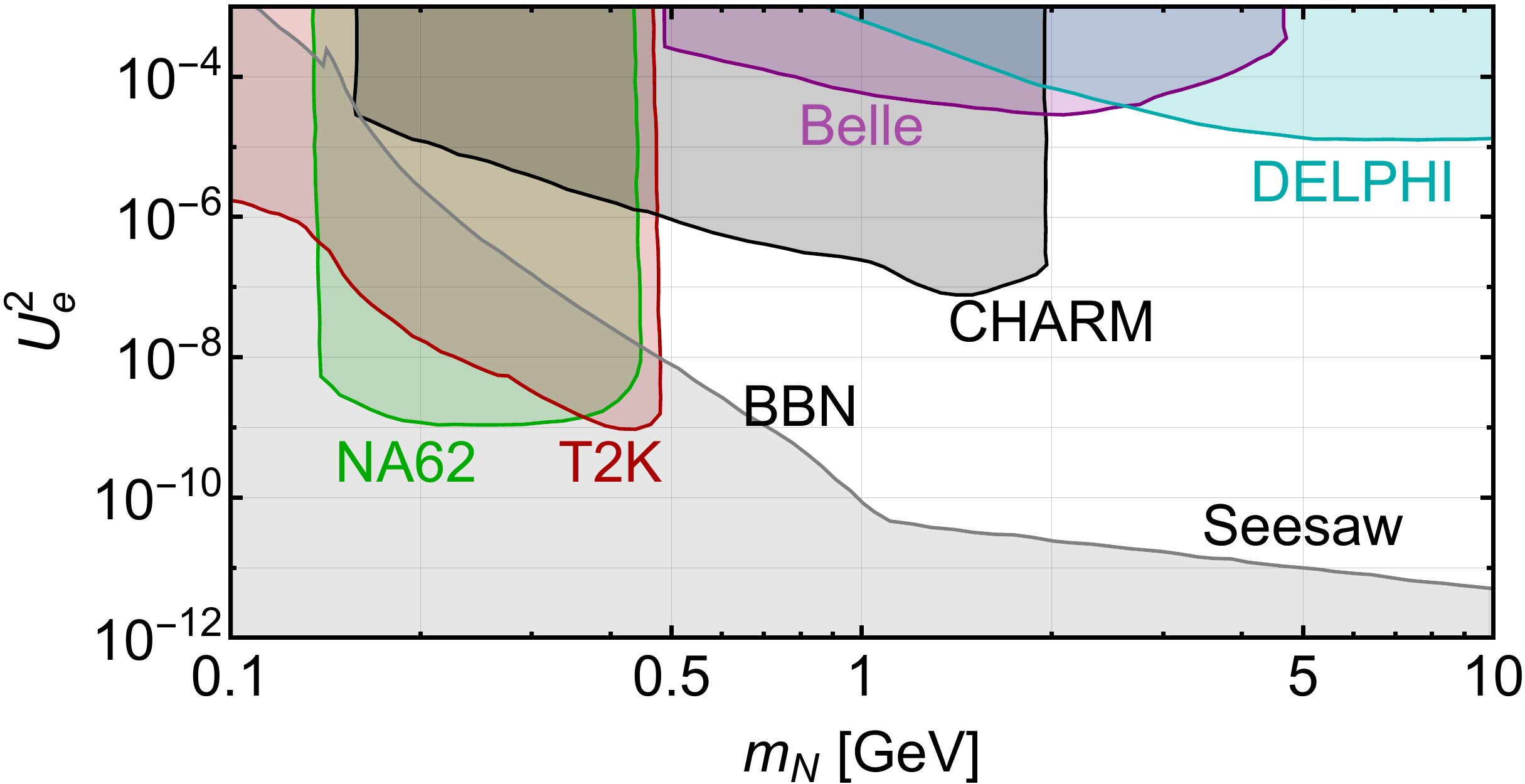}
    
    \includegraphics[width=0.68\textwidth]{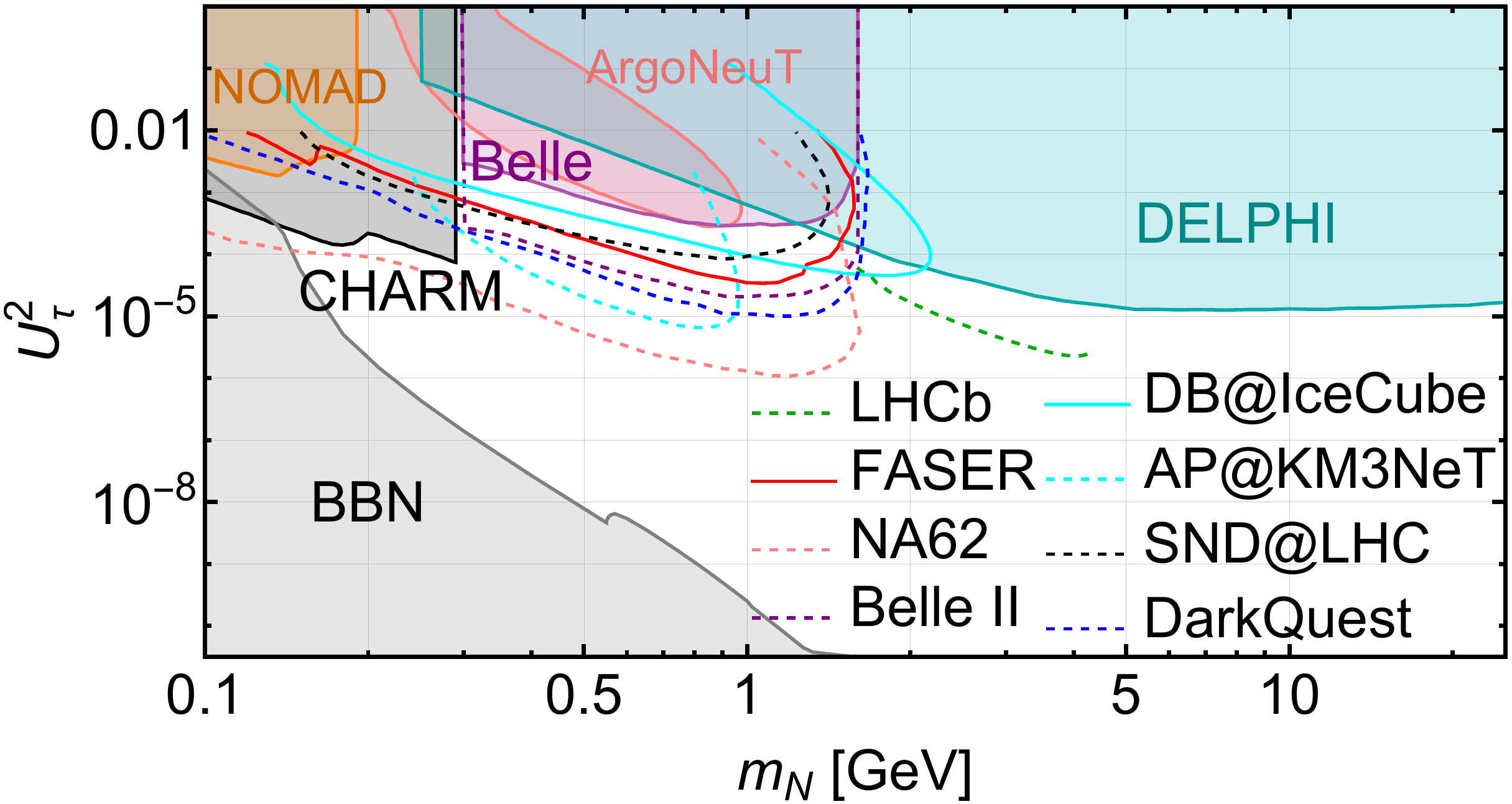}
    \caption{The parameter space of HNLs with the pure $e$ (the left panel) and $\tau$ (the right panel) mixing. The current bounds are from CHARM~\cite{Bergsma:1985is,Orloff:2002de}, NA62~\cite{NA62:2020mcv}, T2K~\cite{T2K:2019jwa}, Belle~\cite{Liventsev:2013zz, Dib:2019tuj},  DELPHI~\cite{Abreu:1996pa}, NOMAD~\cite{NOMAD:2001eyx}, ArgoNeuT~\cite{ArgoNeuT:2021clc}, see also~\cite{Agrawal:2021dbo} for a review and the references therein. For the pure $\tau$ mixing, we do not show the constraints imposed by the T2K experiment (unlike it is done in~\cite{Agrawal:2021dbo}), since they are reported for non-zero couplings $U_{e/\mu}$ which dominate the production\protect\footnotemark[1]. Constraints from the CHARM experiment are taken from the literature, while our re-analysis for them is shown in Fig.~\ref{fig:results}. The light gray domain corresponds to couplings that are either excluded by BBN~\cite{Boyarsky:2020dzc,Sabti:2020yrt} or too small to provide active neutrino masses. For the pure $\tau$ mixing, we also show sensitivities of the next generation Intensity frontier experiments (see text for details). In cyan, we show HNL parameter space that may be probed by neutrino observatories: the solid line shows the sensitivity of IceCube to the ``double bang'' signature from~\cite{Coloma:2017ppo}, while the dashed line corresponds to the sensitivity of KM3NeT to decays of HNLs produced in the atmosphere, see text and Appendix~\ref{app:KM3NeT} for details.}
    \label{fig:old-experiments-sensitivity}
    
    \footnotetext{The T2K experiment~\cite{T2K:2019jwa} is based on a search for decays of HNLs produced from kaons $K\to l_\alpha N$. Such decays can occur only through the $e/\mu$ mixings due to the small mass of kaon $m_K= \unit[493]{MeV}$.}
\end{figure*}

The current bounds on $U^2_e$ and $U^2_\tau$ for HNLs with pure mixing are shown in Fig.~\ref{fig:old-experiments-sensitivity} as reported in~\cite{Agrawal:2021dbo}. In the GeV mass range, the constraints on the mixing angle $U^2_\tau$ are orders of magnitude weaker as compared to the constraints on $U^2_{e/\mu}$ (constraints for the $\mu$ mixing are similar to the ones for the $e$ mixing). 
Namely, for the $e/\mu$ mixing, the large values of the couplings for HNLs with masses $m_{K}\lesssim m_{N}\lesssim m_{D}\simeq \unit[2]{GeV}$ are excluded by the CHARM experiment~\cite{Bergsma:1983rt,Bergsma:1985is}, while for the $\tau$ mixing CHARM constraints on $U_{\tau}$ are reported in the literature only for masses $m_{N}<\unit[290]{MeV}$. 
The reason is the following: the original analysis~\cite{Bergsma:1983rt,Bergsma:1985is} is based on negative results for searches for decays of feebly interacting particles into one of the possible dilepton pair -- $\mu e, \mu\mu, \mu e$. For HNLs, they consider only decays mediated through the charged current (CC) interaction (see Fig.~\ref{fig:hnl-decays}, diagram (a)) that give rise to leptonic decays \begin{equation}
N_{\alpha} \to l_{\alpha}\bar{l}_{\beta}\nu_{\beta}, \quad \beta = e,\mu,\tau
\label{eq:charm-decays-charged}
\end{equation}

If only CC interactions are taken into account, the search is suitable to constrain the mixing of HNLs with $\nu_e$ and $\nu_{\mu}$. To search for CC mediated decays via the $\tau$ mixing (which necessarily include a $\tau$ lepton), the HNL mass should be $m_{N}> m_{\tau}\simeq m_{D}$ in this model. Such HNLs are mainly produced in decays of heavy $B$ mesons, the number of which at CHARM is insufficient to provide enough events for the couplings that are not excluded (see Fig.~\ref{fig:old-experiments-sensitivity}). Therefore, HNLs that mix only with $\nu_{\tau}$ cannot be constrained by CHARM data using only the decays via CC.

In order to constrain the $\tau$ mixing angles of the light HNLs $m_N< m_\tau$, one should include the interactions via the neutral current (NC) into the analysis, see Fig.~\ref{fig:hnl-decays} (diagram (b)). 
In this case, the dileptonic decays are
\begin{equation}
N_{\alpha} \to \nu_{\alpha}l_{\beta}\bar{l}_{\beta},
\label{eq:charm-decays-neutral}
\end{equation}
and do not require the creation of a $\tau$ lepton for the pure $\tau$ mixing.

The works 
\cite{Orloff:2002de,Ruchayskiy:2011aa,Chrzaszcz:2019inj} have re-analyzed the CHARM constraints on HNLs by including also the neutral current processes. However, their analysis was insufficient to put the bounds on the pure $\tau$ mixing in GeV mass range. Namely, the work~\cite{Orloff:2002de} (the results of which are used in~\cite{Agrawal:2021dbo}) has limited the study of the mass range by $m_{N}<\unit[290]{MeV}$, while~\cite{Ruchayskiy:2011aa,Chrzaszcz:2019inj} considered the decays of HNLs via neutral currents but did not include the production of HNLs from $\tau$ lepton (the diagrams (c) and (d) in Fig.~\ref{fig:hnl-prod}). As a result, these works did not report any CHARM limits on the pure $\tau$ mixing.

To conclude, in the GeV mass range of HNLs, there is a gap in the parameter space probed by past experiments, possibly only due to the lack of the analysis (see Fig.~\ref{fig:old-experiments-sensitivity}). To close this gap, the searches for HNLs that mix mainly with $\tau$ neutrinos were considered among scientific goals for several experiments: displaced decays at FASER~\cite{Ariga:2018uku,Beacham:2019nyx}, Belle II~\cite{Dib:2019tuj}, SND@LHC~\cite{Boyarsky:2021moj}, DarkQuest~\cite{Batell:2020vqn}, and NA62 in the dump mode~\cite{Beacham:2019nyx}; prompt decays at LHCb~\cite{Cvetic:2019shl,Chun:2019nwi}; and double bang signature at IceCube, SuperKamiokande, DUNE and HyperKamiokande~\cite{Coloma:2017ppo,Atkinson:2021rnp}.

The planned neutrino observatory KM3NeT~\cite{Adrian-Martinez:2016fdl} working as an atmospheric beam dump may have sensitivity to such HNLs as well. Namely, HNLs may be produced in numerous collisions of cosmic protons with atmospheric particles, then reach the detector volume located deeply underwater in the Mediterranean Sea, and further decay into a dimuon pair inside. Such combination of decay products may be in principle distinguished from the SM events due to neutrino scatterings and penetrating atmospheric muons. We discuss this signature in more detail and estimate the sensitivity of KM3NeT to HNL produced in the atmosphere in Appendix~\ref{app:KM3NeT}, and make the conclusions in Sec.~\ref{sec:sensitivity}. The resulting sensitivity is shown in Fig.~\ref{fig:old-experiments-sensitivity}.

Yet, the main result of this paper is to remove the above-mentioned limitations of the analysis for HNLs interacting only with $\tau$ neutrinos existing in the literature, and demonstrate that the results of the CHARM experiment do imply the constraints on the $\tau$-mixing of HNLs with the masses in the range $\unit[290]{MeV}<m_{N}<\unit[1.6]{GeV}$ that are \emph{two orders of magnitude stronger} than previously reported in the literature. Our results are shown in Fig.~\ref{fig:results}.

The CHARM bounds re-analysis presented in this paper may by similarly applied for the re-analysis of bounds coming from the NOMAD experiment~\cite{NOMAD:2001eyx}. However, due to the smaller intensity of the proton beam at NOMAD and simultaneously similar geometric acceptance of the decay volume, the bounds imposed by NOMAD are sub-dominant, and we therefore do not make the re-analysis in this work.

\section{CHARM experiment}
\label{sec:charm}

\begin{figure}[h!]
    \centering
    \includegraphics[width=\linewidth]{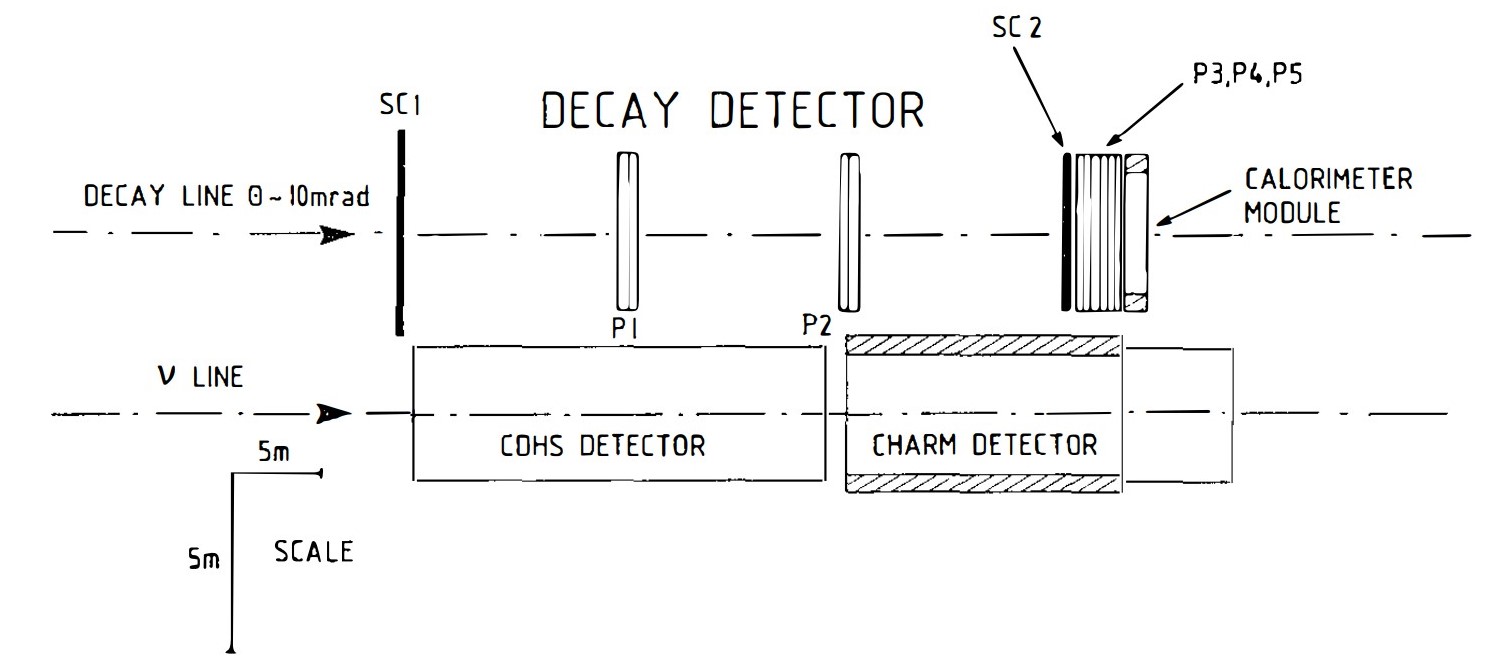}
    \caption{The layout of the CHARM facility, adopted from~\cite{Bergsma:1985is}.}
    \label{fig:charm}
\end{figure}

\begin{figure*}[ht]
    \centering
    \includegraphics[width = \textwidth]{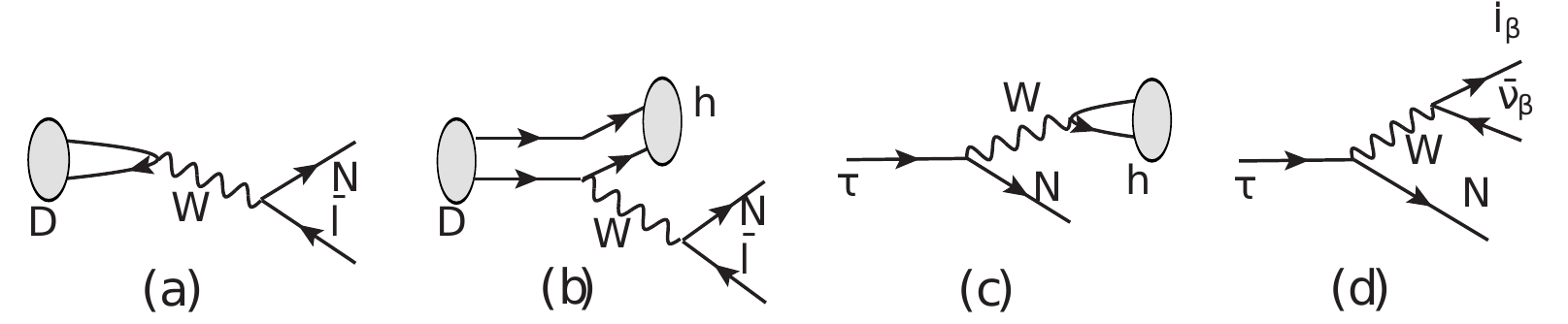}
    \caption{Diagrams of HNL production in leptonic and semileptonic decays of $D$ mesons: $D_{s}, D^{0},D^{\pm}$ (diagrams (a), (b)), and $\tau$ lepton, which is produced in decays of $D_{s}$ meson only (diagrams (c), (d)).}
    \label{fig:hnl-prod}
\end{figure*}

The CHARM experiment~\cite{Bergsma:1983rt, Bergsma:1985is} was a proton beam dump operating at the 400 GeV CERN SPS. The total number of exposed protons was split into $1.7\cdot 10^{18}$ protons on a solid copper target and $0.7\cdot 10^{18}$ on a laminated copper target with the $1/3$ effective density. Searches for decays of HNLs were performed in the $l_{\text{fid}}=\unit[35]{m}$ long decay region (see Fig.~\ref{fig:charm}) defined by the two scintillator planes SC1 and SC2, located at the distance $l_{\text{min}} = \unit[480]{m}$ from the copper target. The decay detector covered the $\unit[3.9 \cdot 10^{-5}]{sr}$ solid angle and had the transverse dimensions $\unit[3\times 3]{m^2}$, with the center displaced by $\unit[5]{m}$ from the axis. The fine-grain calorimeter at CHARM was aimed to detect inelastic scattering of electrons and muons produced in hypothetical decays of HNLs~\cite{Diddens:1980xz}. The sets of tube planes P1-P5~\cite{DIDDENS1980189} were installed to improve the reconstruction of the decay vertex and the angular resolution.

\section{Phenomenology of HNLs at CHARM}
\label{sec:phenomenology}
\subsection{Production}
\label{sec:production}
At the SPS energy of $\unit[400]{GeV}$, HNLs with mass at the GeV scale may be produced directly either in the proton-target collisions, or in the decays of secondary particles: $B,D$ mesons and $\tau$ leptons. The direct HNL production competes with strong interaction processes, while the production from secondary particles -- with weak interactions. As a result, the latter process is dominant even taking into account small production probability of mesons~\cite{Bondarenko:2018ptm}, and the former may be completely neglected. However, similarly to the other experiment operating at SPS -- NA62 in the dump mode, the CHARM experiment has no sensitivity to the HNLs produced from $B$ mesons, implying the lower bound on the probed mass $m_{N} \lesssim m_{D_{s}}\simeq \unit[2]{GeV}$.\footnote{To search for HNLs created in the decays of $B$ mesons at SPS, an experiment like SHiP~\cite{Anelli:2015pba} with significantly larger beam intensity delivered to the experiment and much better geometrical acceptance would be required.}

Let us define the HNL that mixes only with $\nu_{\alpha}$ by $N_{\alpha}$. Neglecting the direct production channels, the total number of $N_{\alpha}$ produced at CHARM is given by:
\begin{multline}
    \mathcal{N}^{(\alpha)}_{\text{prod}} = 2\mathcal{N}_{c\bar{c}}\cdot \big[ \sum_{D_{i}} f_{c\to D_{i}}\text{Br}(D_{i}\to N_{\alpha}X)+ \\ + f_{c\to D_{s}}\cdot \text{Br}(D_{s}\to \tau\bar{\nu}_{\tau})\cdot \text{Br}(\tau \to N_{\alpha}X)\big],
    \label{eq:Nprod}
\end{multline}
with $\mathcal{N}_{c\bar{c}}$ being the total number of quark-antiquark $c\bar{c}$ pairs produced at CHARM, $D_{i}=D^{\pm}$, $D^0$, $D_s$, and $f_{c\to D_{i}}$ the corresponding quark fragmentation fractions at SPS. The first term in the brackets describes the production from decays of $D$ mesons (diagrams (a), (b) in Fig.~\ref{fig:hnl-prod}) and the second -- from $\tau$ leptons in the $D_s\to\tau\to N$ decay chain (diagrams (c), (d) in Fig.~\ref{fig:hnl-prod}). $\text{Br}(D_{i}\to N_{\alpha}X)$, $\text{Br}(\tau \to N_{\alpha}X)$ are the branching ratios. The second term includes a small factor $f_{c\to D_{s}}\cdot \text{Br}(D_{s}\to \tau\bar{\nu}_{\tau}) \simeq 5\cdot 10^{-3}$; for the given HNL mass, it is suppressed as compared to the first term as soon as the production from $D$ is allowed.

The original analysis of the CHARM collaboration~\cite{Bergsma:1983rt,Bergsma:1985is} considered the mixing $\alpha = e,\mu$, for which decays from $D$ mesons are possible for any mass in the range $m_{N}<m_{D_{s}} - m_{l_{\alpha}}\approx \unit[1.9]{GeV}$, and the production from $\tau$ decays may be completely neglected, according to the discussion above. For the $\tau$ mixing, however, the kinematic threshold of the production from $D$, $D_{s}\to \tau + N$, is $m_{D_{s}}-m_{\tau}\approx \unit[190]{MeV}$, and only the second summand in Eq.~\eqref{eq:Nprod} contributes for heavier HNLs. 

\begin{figure*}[t]
    \centering
    \includegraphics[width=0.5\textwidth]{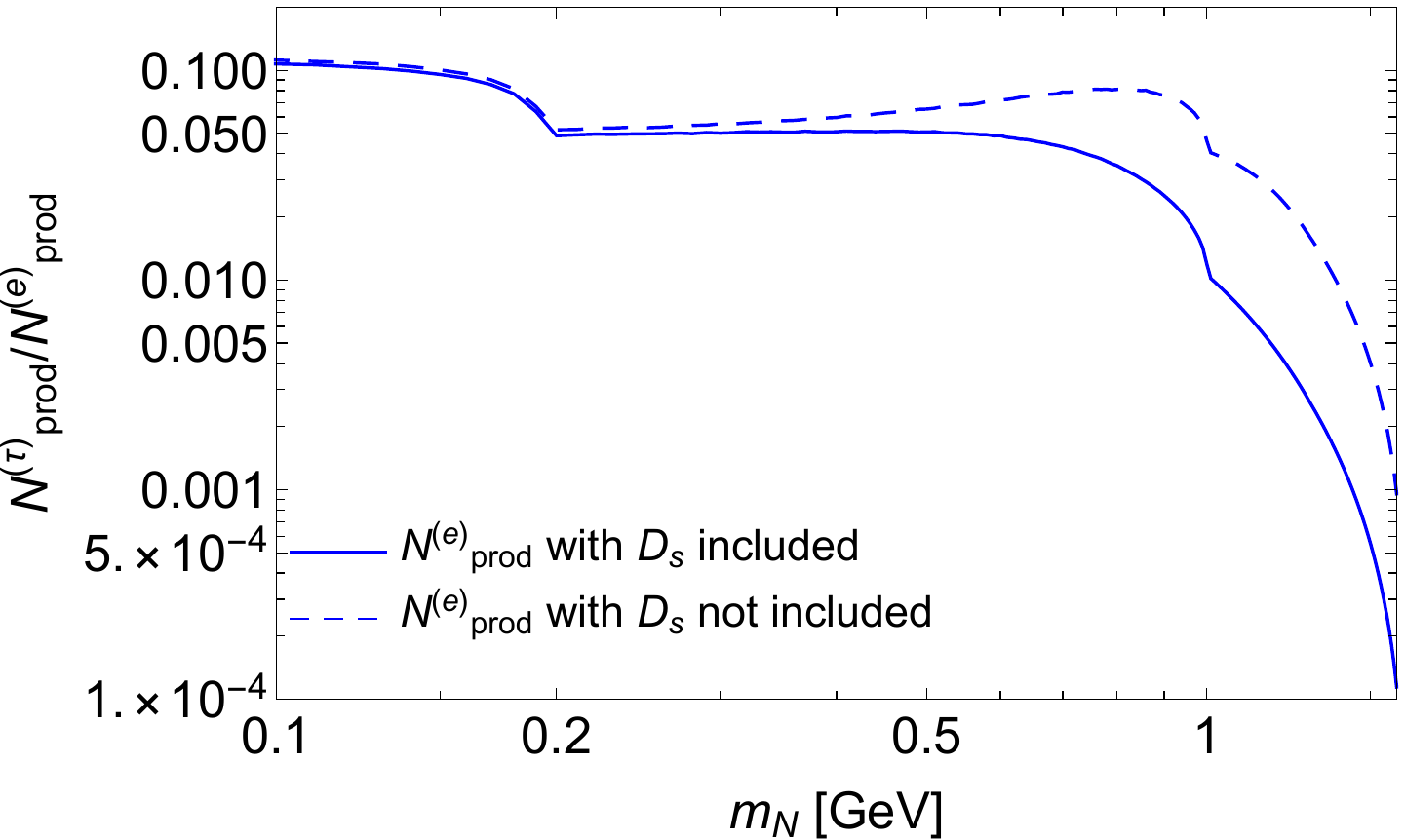}
    \caption{The HNL mass dependence of the ratio of the numbers of produced HNLs with pure $\tau$ and $e$ mixing $N_{\text{prod}}^{(\tau)}/N_{\text{prod}}^{(e)}$, see Eq.~\eqref{eq:Nprod-ratio}, assuming the same values of the mixing angles $U_{e}^{2} = U_{\tau}^{2}$ for the two models. The solid line corresponds to $N_{\text{prod}}^{(e)}$ calculated keeping the production from all $D$ mesons $D^{+},D^{0},D_{s}$, while the dashed line corresponds to the estimate of $N_{\text{prod}}^{(e)} \equiv N_{\text{prod}}^{\text{CHARM}}$ calculated without the contribution of $D_{s}$, as has been done in the analysis~\cite{Bergsma:1985is} by the CHARM collaboration (see text for details). The kinks at $m_{N} = m_{D_{s}} - m_{\tau}\approx \unit[200]{MeV}$ an $m_{N}\approx m_{\tau} - m_{\rho}\approx \unit[1]{GeV}$ correspond to kinematic thresholds of the production channels $D_{s}\to N+\tau$, $\tau \to N+\rho$ correspondingly.}
    \label{fig:prod-ratio}
\end{figure*}

Let us estimate how many HNLs with $\tau$ mixing are produced as compared to those with $e$ mixing. From~\eqref{eq:Nprod}, the ratio $N_{\text{prod}}^{(\tau)}/N_{\text{prod}}^{(e)}$ is
\begin{widetext}
\begin{equation}
    \frac{N^{(\tau)}_{\text{prod}}}{N^{(e)}_{\text{prod}}} = \frac{\sum_{D_{i}} f_{c\to D_{i}}\text{Br}(D_{i}\to N_{\tau}X)+f_{c\to D_{s}}\text{Br}(D_{s}\to \tau\bar{\nu}_{\tau})\text{Br}(\tau\to N_{\tau}X)}{\sum_{D_{i}} f_{c\to D_{i}}\text{Br}(D_{i}\to N_{e}X)},
    \label{eq:Nprod-ratio}
\end{equation}
\end{widetext}
Assuming the same values of mixing angles $U_{e}^{2} = U_{\tau}^{2}$ for the two models with pure $e$/$\tau$ mixing, the ratio $\text{Br}(\tau \to N_\tau X)/\sum f_{c\to D}\text{Br}(D\to N_e X)$ varies in the $1-10$ range for masses $m_N \lesssim \unit[1.3]{GeV}$ and quickly drops at the kinematic threshold $m_N\approx m_\tau$~\cite{Bondarenko:2018ptm}. In particular, for masses $m_{N}\gtrsim \unit[800]{MeV}$, where the dominant contribution to the HNL production with $e$ mixing comes from $D_{s}$, we have
\begin{equation}
     \frac{N^{(\tau)}_{\text{prod}}}{N^{(e)}_{\text{prod}}} \approx \text{Br}(D_{s}\to \tau\bar{\nu}_{\tau})\cdot \frac{\text{Br}(\tau \to N_{\tau}X)}{\text{Br}(D_{s}\to N_{e}X)} <4\cdot 10^{-2}
\end{equation}
The mass dependence of the ratio $N_{\text{prod}}^{(\tau)}/N_{\text{prod}}^{(e)}$ obtained from Eq.~\eqref{eq:Nprod-ratio} is shown in Fig.~\ref{fig:prod-ratio}. 

It is important to note, that in the original analysis~\cite{Bergsma:1985is}, as well as in the re-analyses~\cite{Ruchayskiy:2011aa,Chrzaszcz:2019inj}, the production from $D_{s}$ has not been taken into account for the $e$ mixing. In the mass range $m_{N}\gtrsim \unit[800]{MeV}$, this leads to the underestimate of the number of produced HNLs, $N_{\text{prod}}^{\text{CHARM}}$, by a factor $1/6$ (see Fig.~\ref{fig:hnl-prod}).
\vspace{2cm}

\subsection{Decays and their detection}
\vspace{-3mm}
For a given number of produced HNLs, the number of detected events $N_{\text{events}}^{(\alpha)}$ for the given mixing $\alpha$ depends on
\begin{compactenum}
    \item Geometrical factors -- in order to be detected, produced HNLs need to point in the angular coverage of the CHARM decay volume, decay inside it, and their decay products must then reach the detector and be successfully reconstructed. These factors are: geometrical acceptance $\epsilon_{\text{geom}}$, i.e. the fraction of produced HNLs traveling in the direction of the CHARM detector; the mean HNL gamma factor $\gamma_{N}$; the decay acceptance $\epsilon_{\text{decay}}$, i.e. the fraction of HNL decay products that point to the CHARM detector for HNLs that decay inside the fiducial volume.
    \item The branching ratio $\text{Br}(N_{\alpha}\to l^{+}l'^{-}\nu)$ of the channels $N_{\alpha}\to e^{+}e^{-}\nu, \quad N_{\alpha}\to \mu^{+}\mu^{-}\nu, \quad N_{\alpha}\to e^{-}\mu^{+}\nu$ (and their charge conjugated counterparts) used for detection at CHARM~\cite{Bergsma:1985is}.
\end{compactenum}

The formula for $N_{\text{events}}^{(\alpha)}$ is:

\begin{widetext}
\begin{equation}
    N^{(\alpha)}_{\text{events}} = N_{\text{prod}}^{(\alpha)}\cdot \epsilon^{(\alpha)}_{\text{geom}}\cdot \sum_{l,l' = e,\mu}P_{\text{decay}}^{(\alpha)}\cdot \text{Br}(N_{\alpha}\to ll'\nu)\cdot \epsilon_{\text{det},ll'}\cdot \epsilon^{(\alpha)}_{\text{decay}},
    \label{eq:nevents-main-0}
\end{equation}
\end{widetext}
where $P_{\text{decay}}^{(\alpha)} = e^{-l_{\text{min}}/c\tau^{(\alpha)}_{N}\gamma^{(\alpha)}_{N}}-e^{-(l_{\text{min}}+l_{\text{fid}})/c\tau^{(\alpha)}_{N}\gamma^{(\alpha)}_{N}}$ is the decay probability, and $\epsilon_{\text{det},ll'}$ is the reconstruction efficiency for the given channel.

We will see below that geometrical factors are the same for $e,\mu$ and $\tau$ mixing, while the branching ratio is smaller for the $\tau$ mixing channels, as in the former case both decays via the charged and neutral currents are relevant, while in the latter only the neutral current contribute.

Let us start by considering the lower bound of the sensitivity of the CHARM experiment, i.e. the minimal mixing angles that it may probe (the upper bound will be discussed in Sec.~\ref{sec:sensitivity}).
In this regime, the decay length of the HNL $c\tau^{(\alpha)}_{N}\gamma^{(\alpha)}_{N}$ is much larger than the geometric scale of the experiment, $c\tau^{(\alpha)}_{N}\gamma^{(\alpha)}_{N}\gg l_{\text{min}}+l_{\text{fid}} \approx \unit[515]{m}$. Then $P_{\text{decay}}^{(\alpha)} \approx \frac{l_{\text{fid}}}{c\gamma^{(\alpha)}_{N}}\cdot \Gamma(N_{\alpha})$, where $\Gamma(N_{\alpha})$ is the total decay width, and it is convenient to rewrite Eq.~\eqref{eq:nevents-main-0} in the form
\begin{widetext}
\begin{equation}
    N^{(\alpha)}_{\text{events}} \approx N_{\text{prod}}^{(\alpha)}\times \epsilon^{(\alpha)}_{\text{geom}}\cdot \sum_{l,l' = e,\mu}\frac{l_{\text{fid}}}{c\gamma^{(\alpha)}_{N}}\cdot \Gamma(N_{\alpha}\to ll'\nu)\epsilon_{\text{det},ll'}\cdot \epsilon^{(\alpha)}_{\text{decay}},
    \label{eq:nevents-main}
\end{equation}
\end{widetext}
where $\Gamma(N_{\alpha}\to l^{+}l'^{-}\nu)$ is the decay width into the dilepton pair $ll'$.

We will first discuss the difference in  $\Gamma(N_{\alpha}\to l^{+}l'^{-}\nu)$ between the cases of $e$ and $\tau$ mixings. Decays into dileptons occur via charged and neutral current, see Fig.~\ref{fig:hnl-decays}. For the NC mediated processes, the kinematic threshold $m_{N} > 2m_e \approx \unit[1]{MeV}$ is mixing-independent. In contrast, for the CC mediated process for the $\tau$ mixing this threshold is $m_{N} > m_{\tau}+m_{e} \approx \unit[1.77]{GeV}$, and HNLs lighter than $\tau$ lepton may decay into dileptons only via NC. 

Decay widths for the processes $N_{\alpha}\to l^{+}l'^{-}\nu$ for $m_N\gg m_l+m_{l'}$ may be given in the unified form 
\begin{equation}
\Gamma(N_\alpha \to l^{+} l'^{-} \nu) = c^{(\alpha)}_{ll'\nu}\frac{G_F^2 m_N^5}{192\pi^3},
\label{eq:dileptonic-width}
\end{equation}
where the coefficients $c^{(\alpha)}_{ll'\nu}$ are given in Table~\ref{tab:NCCCcoef}~\cite{Bondarenko:2018ptm}. For $N_{e}$, the largest decay width is $\Gamma(N_{e}\to \mu^{+} e^{-}\nu_{\mu})$, where only CC contributes. The width $\Gamma(N_{e}\to e^{+}e^{-}\nu_{e})$ is smaller: 
\begin{equation}
 \Gamma(N_{e}\to e^{-}e^{+}\nu_{e})/\Gamma(N_{e}\to e^{-}\mu^{+}\nu_{\mu})\approx 0.59,  
\end{equation}
because both NC and CC contribute in this process and interfere destructively. The smallest width is $\Gamma(N_{e}\to \mu^{+}\mu^{-}\nu_{e})$, with the process occurring only via NC. For $N_{\tau}$, there is no process $N_{\tau} \to e\mu \nu$, while in the process $N_{\tau}\to e^{+}e^{-}\nu_{\tau}$ only NC contributes, and thus the width is smaller than for $N_{e}$: 
\begin{equation}
\Gamma(N_{\tau}\to e^{+}e^{-}\nu_{\tau})/\Gamma(N_{e}\to e^{+}e^{-}\nu_{e}) \approx 0.22
\end{equation}
For the decay into a dimuon pair, we have $\Gamma(N_{\tau}\to \mu^{+}\mu^{-}\nu_{\tau}) = \Gamma(N_{e}\to \mu^{+}\mu^{-}\nu_{e})$.

As a result, for $m_{N}\gg m_{\mu}$ the ratio of the factors $\sum_{l,l'}\Gamma(N_{\alpha}\to ll'\nu)\epsilon_{\text{det},ll'}$ entering Eq.~\eqref{eq:nevents-main} is given by
\begin{equation}
    \frac{\sum_{l}\Gamma(N_{\tau}\to ll)\epsilon_{\text{det},ll}}{\sum_{l,l'}\Gamma(N_{e}\to ll')\epsilon_{\text{det},ll'}} \approx 0.16
    \label{eq:decay-widths-ratio}
\end{equation}
Here and below, we use the values of the efficiencies $\epsilon_{\text{det},ll'}$ as reported in~\cite{Bergsma:1985is} for the HNL mass $m_{N} = \unit[1]{GeV}$: $\epsilon_{\text{det},ee} = 0.6$, $\epsilon_{\text{det},e\mu} = 0.65$, $\epsilon_{\text{det},\mu\mu} = 0.75$.

In the original analysis for the $e$ mixing by the CHARM collaboration~\cite{Bergsma:1983rt,Bergsma:1985is}, the Dirac nature of HNLs has been assumed (the decay widths are twice smaller), and only the CC interactions have been considered. Instead of Eq.~\eqref{eq:decay-widths-ratio}, the ratio becomes
\begin{equation}
\frac{2\sum_{l}\Gamma(N_{\tau}\to ll)\epsilon_{\text{det},ll}}{\sum_{l,l'}\Gamma_{\text{CC}}(N_{e}\to ll')\epsilon_{\text{det},ll'}} \approx 0.27
    \label{eq:decay-widths-ratio-CHARM}
\end{equation}
 
\begin{table}[h!]
     \centering
     \begin{tabular}{|c|c|c|}
          \hline
          Process & $c^{(\alpha)}_{ll'\nu}$  \\
          \hline
          $N_{e/\tau} \to \mu^{+}\mu^{-}\nu_{e/\tau}$& $\frac{1}{4}(1-4\sin^2 \theta_W + 8\sin^4 \theta_W)\approx 0.13$ \\
          \hline
          $N_{\tau} \to e^{+}e^{-}\nu_{\tau}$& $\frac{1}{4}(1-4\sin^2 \theta_W + 8\sin^4 \theta_W)\approx 0.13$ \\
          \hline
          $N_{e}\to e^{-}\mu^{+}\nu_{\mu}$& $1$ \\ \hline
          $N_{e}\to e^{+}e^{-}\nu_{e}$ & $\frac{1}{4}(1+4\sin^2 \theta_W + 8\sin^4 \theta_W)\approx 0.59$ \\
          \hline
          $N_{e}\to e^{+}e^{-}\nu_{e}$ (CC) & 1 \\
          \hline
     \end{tabular}
     \caption{The values of $c_{ll'\nu}^{(\alpha)}$ in Eq.~\eqref{eq:dileptonic-width} for different decay processes. For the process $N_{e}\to e^{+}e^{-}\nu_{e}$, we also provide the value obtained if including the charged current (CC) contribution only -- the assumption used in~\cite{Bergsma:1985is}.}
     \label{tab:NCCCcoef}
 \end{table} 
 
Let us now discuss geometric factors $\epsilon_{\text{geom}},\gamma_{N},\epsilon_{\text{decay}}$. It turns out that they depend on the mixing pattern weakly, and as a result the geometry does not influence the relative yield of events for $e$ and $\tau$ mixing. Indeed, as was mentioned in Sec.~\ref{sec:production}, HNLs with $\tau$ mixing are produced in decays of $\tau$ leptons, that originate from decays of $D_{s}$. Since $m_{\tau} \simeq m_{D_{s}}$, the angle-energy distribution of $\tau$ leptons is the same as of $D_{s}$ (and hence also other $D$ mesons), whose decays produce HNLs with $e$ mixing. The kinematics of the HNL production from $D$ and $\tau$ is similar: two-body decays (a), (c) and three-body decays (b), (d) in Fig.~\ref{fig:hnl-prod} differ mainly be the replacement a neutrino or a lepton with a hadron $h = \pi, K$. However, since $m_{h}\ll m_{\tau,D}$, the replacement does not lead to the difference in the distribution of produced HNLs. In addition, heavy HNLs with masses $m_{N}\simeq \unit[1]{GeV}$ share the same distribution as their mother particles, and any difference disappear. Therefore, the values $\epsilon_{\text{geom}}$, $\gamma_{N}$ for different mixing are the same with good precision. Next, HNL decays contain the same final states independently of the mixing, and $\epsilon_{\text{decay}}$ can also be considered the same. 

To summarize, the ratio $N_{\text{events}}^{(\tau)}/N_{\text{events}}^{(e)}$ is determined only by the difference in phenomenological parameters -- $N_{\text{prod}}^{(\alpha)}$ and $\Gamma(N_{\alpha}\to ll'\nu)$:
\begin{equation}
    \frac{N_{\text{events}}^{(\tau)}}{N_{\text{events}}^{(e)}} \simeq \frac{N_{\text{prod}}^{(\tau)}}{N_{\text{prod}}^{(e)}}\times \frac{\sum_{l}\Gamma(N_{\tau}\to ll\nu)\epsilon_{\text{det},ll}}{\sum_{l,l'}\Gamma(N_{e}\to ll'\nu)\epsilon_{\text{det},ll'}}
    \label{eq:chi-ratio}
\end{equation}
To compare with the estimate of the number of events for the $e$ mixing made by the CHARM collaboration in~\cite{Bergsma:1985is}, $N_{\text{events}}^{\text{CHARM}}$, we need to take into account their assumptions on the description of HNL production and decays (see the discussion around Eqs.~\eqref{eq:Nprod-ratio} and~\eqref{eq:decay-widths-ratio-CHARM}). The resulting ratio is
\begin{equation}
        \frac{N_{\text{events}}^{(\tau)}}{N_{\text{events}}^{\text{CHARM}}} \simeq \frac{N_{\text{prod}}^{(\tau)}}{N_{\text{prod}}^{\text{CHARM}}}\cdot \frac{2\sum_{l}\Gamma(N_{\tau}\to ll\nu)\epsilon_{\text{det},ll}}{\sum_{l,l'}\Gamma_{\text{CC}}(N_{e}\to ll'\nu)\epsilon_{\text{det},ll'}}
        \label{eq:Nevents-ratio-CHARM}
\end{equation}
 
\section{Results}
\label{sec:sensitivity}

Let us now derive the CHARM sensitivity to the $\tau$ mixing. In~\cite{Bergsma:1985is}, it has been shown that the dilepton decay signature at CHARM is background free. Therefore,  $90\%$ CL  sensitivity  to each mixing is given by the condition
\begin{equation}
    N^{(e,\mu,\tau)}_{\text{events}} > 2.3
    \label{eq:sensitivity}
\end{equation}
Let us define $U^{2}_{\text{lower,\text{CHARM}}}$ as the smallest mixing angle for which the condition~\eqref{eq:sensitivity} is satisfied for the assumptions of the original analysis of~\cite{Bergsma:1985is} (see the discussion above Eq.~\eqref{eq:Nevents-ratio-CHARM}).
As the number of detected events at the lower bound $N^{(\alpha)}_{\text{events}}$ scales with the mixing angle as $N^{(\alpha)}_{\text{events}}\propto U_{\alpha}^{4}$ (where $U_{\alpha}^{2}$ comes from the production and another $U^{2}_{\alpha}$ from decay probability), we can use Eqs.~\eqref{eq:Nevents-ratio-CHARM} and~\eqref{eq:Nprod-ratio} to obtain the lower bound of the sensitivity to the $\tau$ mixing, $U^{2}_{\tau,\text{lower}}$, by rescaling the results reported in~\cite{Bergsma:1985is}: 
\begin{multline}
    \frac{U^{4}_{\tau,\text{lower}}}{U^{4 \text{ CHARM}}_{\text{lower}}} \simeq\\\simeq \frac{N_{\text{prod}}^{\text{CHARM}}}{N_{\text{prod}}^{(\tau)}}\cdot \frac{\sum_{l,l'}\Gamma_{\text{CC}}(N_{e}\to ll'\nu)\epsilon_{\text{det},ll'}}{\sum_{l}\Gamma(N_{\tau}\to l\bar{l}\nu)\epsilon_{\text{det},ll}}\bigg|_{U_{e} = U_{\tau}}.
    \label{eq:rescaling}
\end{multline}
Using the ratio $N_{\text{prod}}^{\text{CHARM}}/N_{\text{prod}}^{(\tau)}$ from Eq.~\eqref{eq:Nprod-ratio} (see also Fig.~\ref{fig:prod-ratio}), and the ratio of decay widths from Eq.~\eqref{eq:decay-widths-ratio-CHARM}, we conclude that in the mass range $m_{N}>\unit[200]{MeV}$ the lower bound for the $\tau$ mixing is a factor $10-100$ weaker than the lower bound for the $e$ mixing reported in~\cite{Bergsma:1985is}. In the domain $m_{D_s}-m_\tau<m_{N} < \unit[290]{MeV}$, we validate the rescaled bound~\eqref{eq:rescaling} by comparing it with the CHARM sensitivity to the $\tau$ mixing from~\cite{Orloff:2002de}, see Appendix~\ref{app:decay-events-number}. 

At the \emph{upper bound of the sensitivity}, the dependence of the number of events on $U^{2}_{\alpha}$ is complicated and the sensitivity cannot be obtained by rescaling the results of~\cite{Bergsma:1985is}. Therefore, we independently compute the number of decay events at CHARM for HNLs with $e$ and $\tau$ mixing and then calculate the sensitivity numerically using Eq.~\eqref{eq:sensitivity}, see Appendix~\ref{app:decay-events-number}. In order to validate this estimate, we compare the resulting sensitivity for the $\tau$ mixing with the rescaled bound~\eqref{eq:rescaling}, and find that they are in very good agreement (Fig.~\ref{fig:epsilons}). Also, we compare our estimate for the $e$ mixing with the CHARM sensitivity to the $e$ mixing from~\cite{Bergsma:1985is}. In our estimates, we include neutral current interactions, the production from $D_{s}$ mesons, and assume that HNLs are Majorana particles. Due to these reasons, we find that for small mixing angles $U_{e}$ and above $m_{N}\gtrsim \unit[1]{GeV}$, the bound imposed by CHARM may be actually improved by up to a factor $3-4$.
\begin{figure*}[ht]
    \centering
    \includegraphics[width=0.65\textwidth]{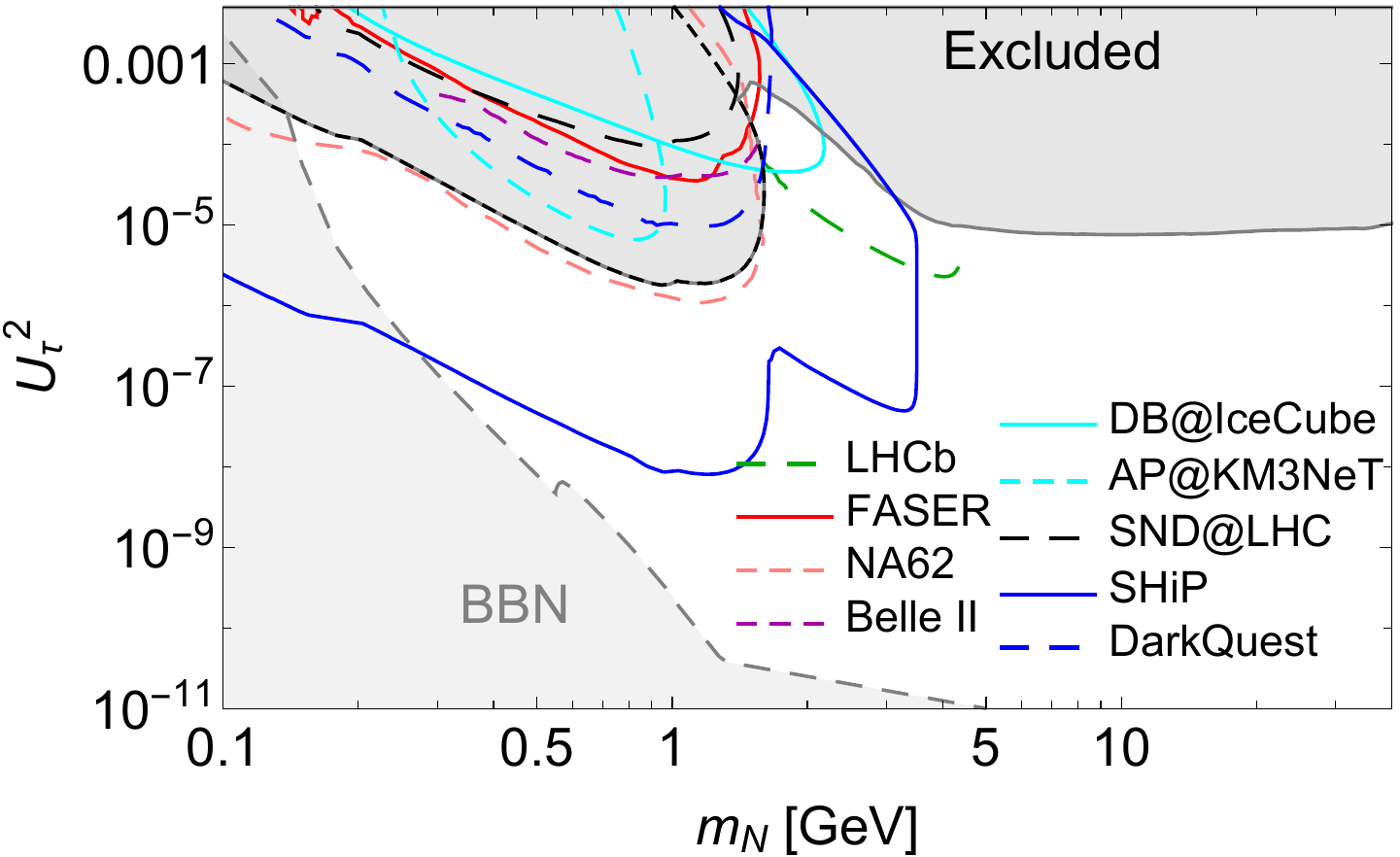}
    
    \includegraphics[width=0.65\textwidth]{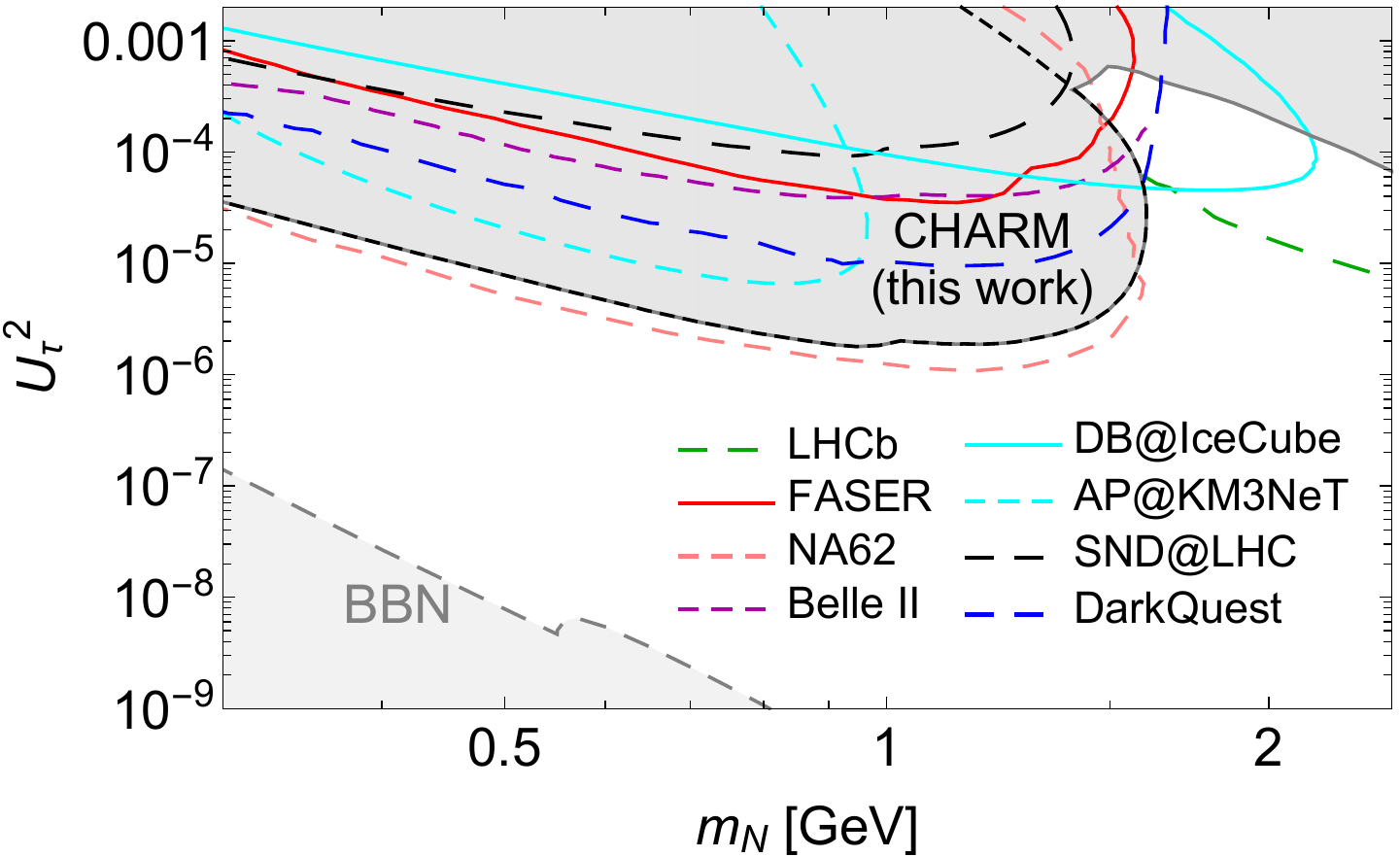}
    \caption{Parameter space of a single Majorana HNL that mixes with $\nu_{\tau}$. The excluded region is a combined reach of the DELPHI~\cite{Abreu:1996pa} and CHARM experiments (our result), which is also highlighted by the short-dashed black line. Bounds from BBN are reproduced from~\cite{Boyarsky:2020dzc,Sabti:2020yrt}. The sensitivity of future experiments is also shown (see text around Fig.~\ref{fig:old-experiments-sensitivity} for details). The top panel covers the HNL mass region $m_{N} = \unit[0.1-35]{GeV}$, while the bottom panel is a zoom-in of the mass domain $m_{N} = \mathcal{O}(\unit[1]{GeV})$.}
    \label{fig:results}
\end{figure*}
Let us comment on errors of our estimates. We used the values of reconstruction efficiencies $\epsilon_{\text{rec},ll}$ reported in~\cite{Bergsma:1985is} for the HNL mass $m_{N} = \unit[1]{GeV}$. Hence, the calculation may be further refined by including HNL mass dependent reconstruction efficiencies. However, as the study~\cite{Orloff:2002de} performed for the $\tau$ mixing and masses $m_{N}<\unit[290]{MeV}$ has shown similar efficiency, we do not expect any significant changes.

Our final results for the $\tau$ mixing are given in Fig.~\ref{fig:results}, where we show the domain excluded by previous experiments together with updated CHARM bounds, and the sensitivity of the future experiments mentioned in Sec.~\ref{sec:introduction}, together with SHiP~\cite{SHiP:2018xqw}. Comparing with Fig.~\ref{fig:old-experiments-sensitivity}, we find that in the mass range $\unit[290]{MeV}<m_{N}<\unit[1.6]{GeV}$ our results improve previously reported bounds on the mixing angle $U^{2}_{\tau}$ by two orders of magnitude. In particular, it excludes a large part of the parameter space that was suggested to be probed by the future experiments.
For instance, Belle II, FASER, DarkQuest and the double bang signature at IceCube have sensitivity only in the narrow domain above the CHARM upper bound, while NA62 may slightly push probed angles to lower values. The same is the case for the sensitivity of KM3NeT in the regime of the atmospheric beam dump, which we estimate in Appendix~\ref{app:KM3NeT}. Significant progress in testing 
the mixing of HNLs with $\nu_{\tau}$ can be achieved by 
LHCb, which probes the complementary mass range $m_{N}>\unit[2]{GeV}$, and dedicated intensity frontier experiments, with SHiP being optimal for searches of HNLs from decays of $D$ mesons and $\tau$ leptons.

\section*{Acknowledgements}
We thank Y.~Cheipesh for careful reading and improving the quality of the manuscript. This project has received funding from the European Research Council (ERC) under the European Union's Horizon 2020 research and innovation programme (GA 694896) and from the NWO Physics Vrij Programme ``The Hidden Universe of Weakly Interacting Particles'', nr. 680.92.18.03, which is partly financed by the Dutch Research Council NWO.

\onecolumngrid

\appendix

\section{CHARM sensitivity based on number of decay events estimate}
\label{app:decay-events-number}
The number of decay events for the pure $\alpha$ mixing at CHARM is given by the formula
\begin{multline}
            N^{(\alpha)}_{\text{events}} = \sum_{X = D,\tau}N_{X}\cdot \text{Br}(X\to N_{\alpha})\times \\ \times \int dEd \theta dz\cdot f_{N_{\alpha}}^{X}(E,\theta)\frac{e^{-l(z)/c\tau_{N}\gamma}}{c\tau^{(\alpha)}_{N}\gamma}\frac{\Delta \phi(\theta,z)}{2\pi} \cdot \epsilon_{\text{decay}}(\theta,z,E)\cdot \text{Br}(N_{\alpha}\to l\bar{l}')\cdot \epsilon_{\text{det},ll'}
            \label{eq:nevents}
        \end{multline}
Here,
\begin{equation}
N_{D_{i}} =N_{\text{PoT}}\times \chi_{c\bar{c}}\times f_{c\to D_{i}}, 
\quad N_{\tau} = N_{D_{s}}\times \text{Br}(D_{s}\to \tau\bar{\nu}_{\tau})
\end{equation}
are the total numbers of $D$ mesons ($D_{i} = D_{s},D^{+},D^{0}$) and $\tau$ leptons, with $N_{\text{PoT}} = 2.4\cdot 10^{18}$ being the total number of proton-target collisions at CHARM and $\chi_{c\bar{c}} \approx 4\cdot 10^{-3}$ the production fraction of the $c\bar{c}$ at SPS energies for a thick target~\cite{CERN-SHiP-NOTE-2015-009}. $\text{Br}_{D_{s}\to \tau} \approx 5.43\%$~\cite{Tanabashi:2018oca} and $f_{c\to D_{i}}$ are given from~\cite{SHiP:2018xqw}. $f_{N_{\alpha}}^{X}$ is the distribution of HNLs produced in decays of $X$ particles in polar angle and energy. $z \in \unit[(480,515)]{m}$ is the longitudinal distance, $\theta \in (3.5/515, 6.5/515)$ is the polar angle coverage of the end of the CHARM's decay volume, while $\Delta \phi(\theta)/2\pi$ is the azimuthal acceptance for HNLs decaying inside the decay volume. $\epsilon_{\text{decay}}$ is the decay acceptance -- a fraction of decay products of HNLs that both point to the detector. Finally, $\epsilon_{\text{det},ll'}$ are reconstruction efficiencies for leptonic decays: $\epsilon_{ee} \approx 60\%$, $\epsilon_{\mu\mu} \approx 75\%$, and $\epsilon_{e\mu} \approx 65\%$, which we use from~\cite{Bergsma:1985is}.

Computing of $f_{N_{\alpha}}(E,\theta)$ requires knowing the distribution of $D$ mesons and $\tau$ leptons $f_{\tau}(E,\theta)$ produced at the CHARM target. We approximate $f_{\tau}$ by the distribution of $D_{s}$ mesons, while for the distribution of $D$ mesons we use FairShip simulations~\cite{CERN-SHiP-NOTE-2015-009} for collisions of the SPS proton beam with a thick Tungsten target.\footnotemark

\footnotetext{ \begin{minipage}{\textwidth}
Although at CHARM the target material is different, we believe that it is still a reasonable approximation.
\end{minipage}
} 

The distribution of HNLs $f_{N_{\alpha}}^{X}(E,\theta)$ has been obtained from $f_{X}(E,\theta)$ semi-analytically using the method from~\cite{Boiarska:2019vid}. 

\begin{figure}[h!]
    \centering
    \includegraphics[width = 0.33\textwidth]{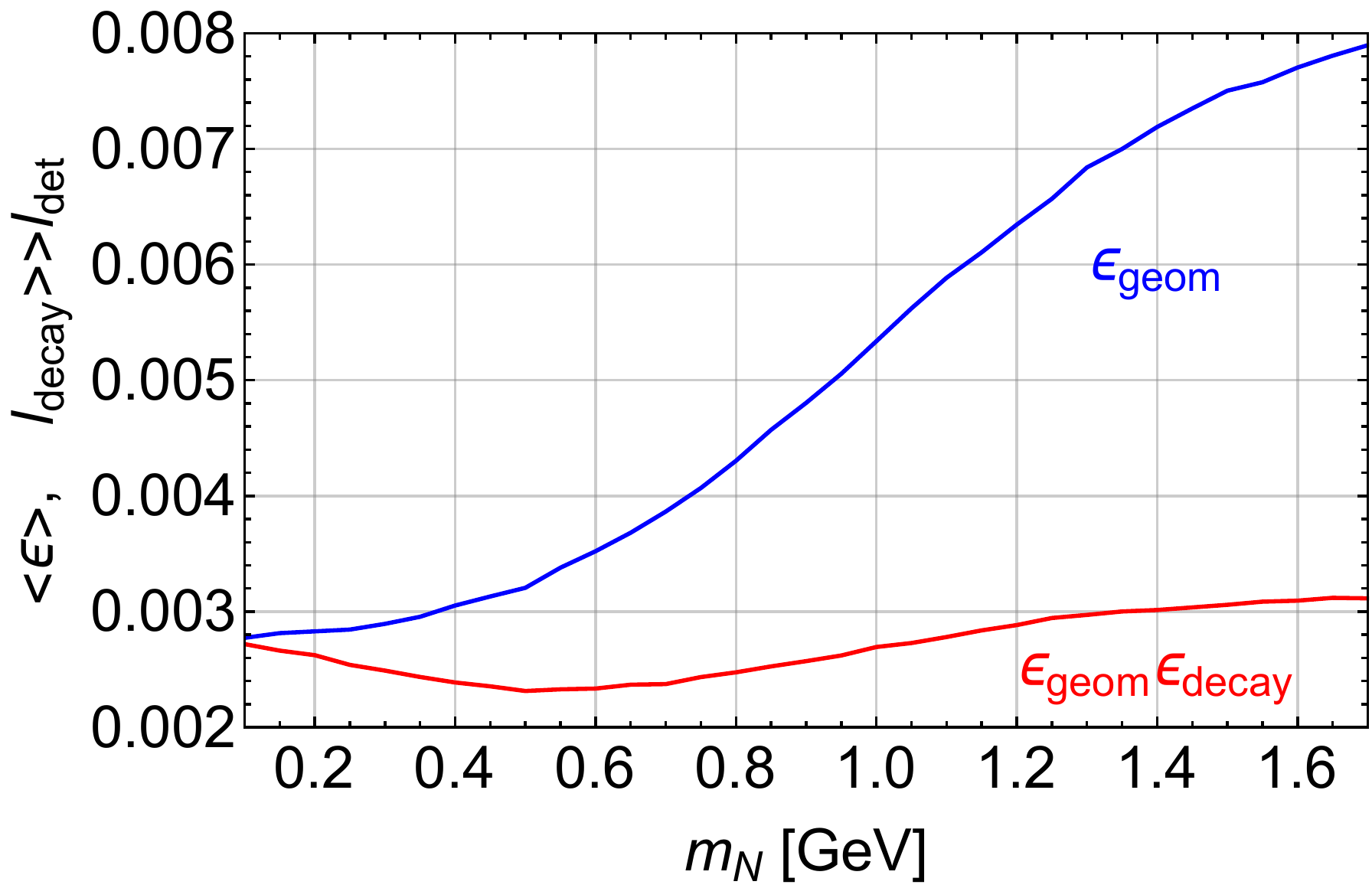}~\includegraphics[width = 0.33\textwidth]{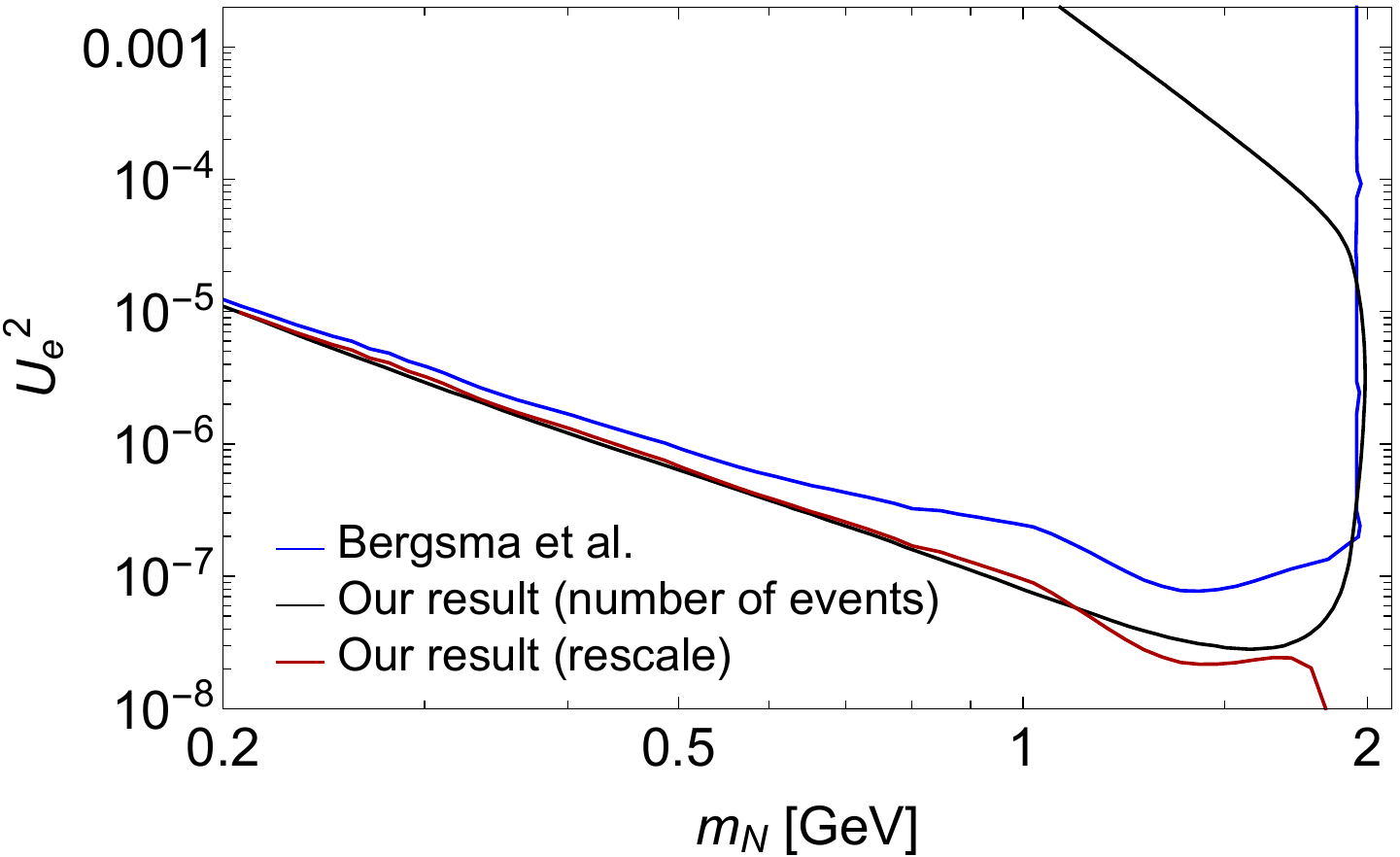}~\includegraphics[width = 0.33\textwidth]{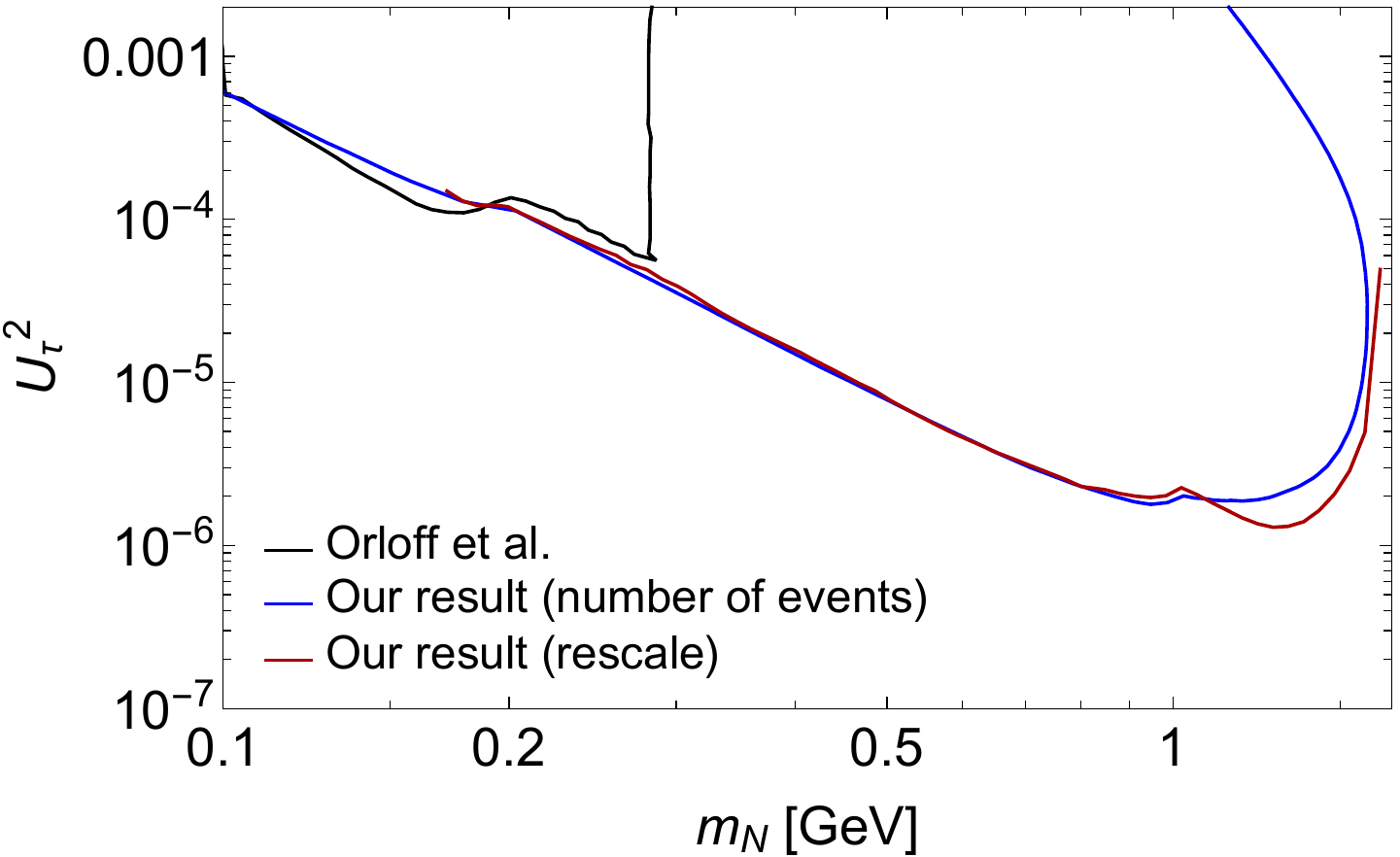}
    \caption{The left panel: fraction of HNLs that point towards the detector (blue line) and fraction of HNLs whose decay products point towards the detector (red line). The middle and right panels: comparison of our estimates of the constraint from the CHARM experiment on the pure $e$ (the middle panel) and $\tau$ mixing (the right panel), with bounds reported in~\cite{Bergsma:1985is} and~\cite{Orloff:2002de}. We show two estimates: the red line corresponds to the rescale of the bound on the $e$ mixing from~\cite{Bergsma:1985is} (see Sec.~\ref{sec:sensitivity} for details), while the blue line is our independent estimate based on Eq.~\eqref{eq:nevents}.}
    \label{fig:epsilons}
\end{figure}

We have estimated $\epsilon_{\text{decay}}$ by using a toy simulation for decays of HNLs inside the decay volume into three massless particles, and requiring the momenta of the two charged leptons to point towards the end of the decay volume. The acceptances are shown in Fig.\ref{fig:epsilons}. 

In order to obtain the excluded domain, we assume the absence of background and require $N_{\text{events}}>2.3$, which corresponds to the 90\% C.L.

The comparison of this estimate with the rescale from Sec.~\ref{sec:sensitivity} and~\cite{Orloff:2002de} is shown in Fig.~\ref{fig:epsilons}. We find that the estimates are in very good agreement. We also show our estimate of the CHARM bounds on the $e$ mixing, which differs from the bounds obtained from~\cite{Bergsma:1985is} by including the production from $D_{s}$ mesons, which dominates masses $m_{N}\gtrsim \unit[700]{MeV}$ (see also Fig.~\ref{fig:prod-ratio}). The resulting sensitivity at the lower bound improves by up to a factor $3-4$ for this mass region. 

\section{Sensitivity of KM3NeT to HNLs}
\label{app:KM3NeT}
In this Appendix, we demonstrate the potential of the neutrino telescope KM3NeT~\cite{Adrian-Martinez:2016fdl} to probe the parameter space of HNLs. 

HNLs may be numerously produced in decays of $\tau$ leptons, originated from the collisions of high-energy cosmic protons with the well-known spectrum
\begin{equation}
      \frac{d\Phi}{d\Omega dtdS dE_{p}} \approx \begin{cases} 1.7 \ E_{p,\text{ GeV}}^{-2.7}\text{ GeV}^{-1}\text{sr}^{-1}\text{cm}^{-2}\text{s}^{-1} , \quad E_{p} < \unit[5\cdot 10^{6}]{GeV} \\ 174 \  E_{p,\text{ GeV}}^{-3}\text{ GeV}^{-1}\text{sr}^{-1}\text{cm}^{-2}\text{s}^{-1} , \quad E_{p} \geqslant \unit[5\cdot 10^{6}]{GeV}
      \end{cases}  
  \end{equation} 
with atmospheric particles. If having significantly large lifetimes, produced HNLs may enter the detector volume of KM3NeT located deep in the Mediterranean Sea and decay there. 

In order to probe the parameter space of HNLs, it is necessary to distinguish their decays from interactions of SM particles also produced in the atmosphere: neutrinos and muons. KM3NeT may only distinguish two event types: track-like, which corresponds to muons penetrating through the detector volume, and cascade-like, which originates from other particles such as electrons and hadrons. Scatterings of neutrinos inside the detector volume produce cascade-like (if no high-energy muons are produced) or combined cascade-like + track-like signature (if high-energy muons are produced), while penetrating atmospheric muons give rise to track-like signature. A possible way to distinguish these events from HNLs is to look for their decays into a di-muon pair, $N\to \mu\bar{\mu}\nu_{\tau}$. They produce a signature of two tracks originated from one point inside the detector volume. Detectors of KM3NeT have energy and angular resolution sufficient precise for resolving the two tracks down to energies of $\simeq \unit[10]{GeV}$~\cite{KM3Net:2016zxf}. Therefore, we believe that the dimuon signature may be reconstructed in the background free regime with high efficiency.

\subsection{Analytic estimates: comparison with CHARM}
\begin{figure}[!h]
    \centering
    \includegraphics[width=0.6\textwidth]{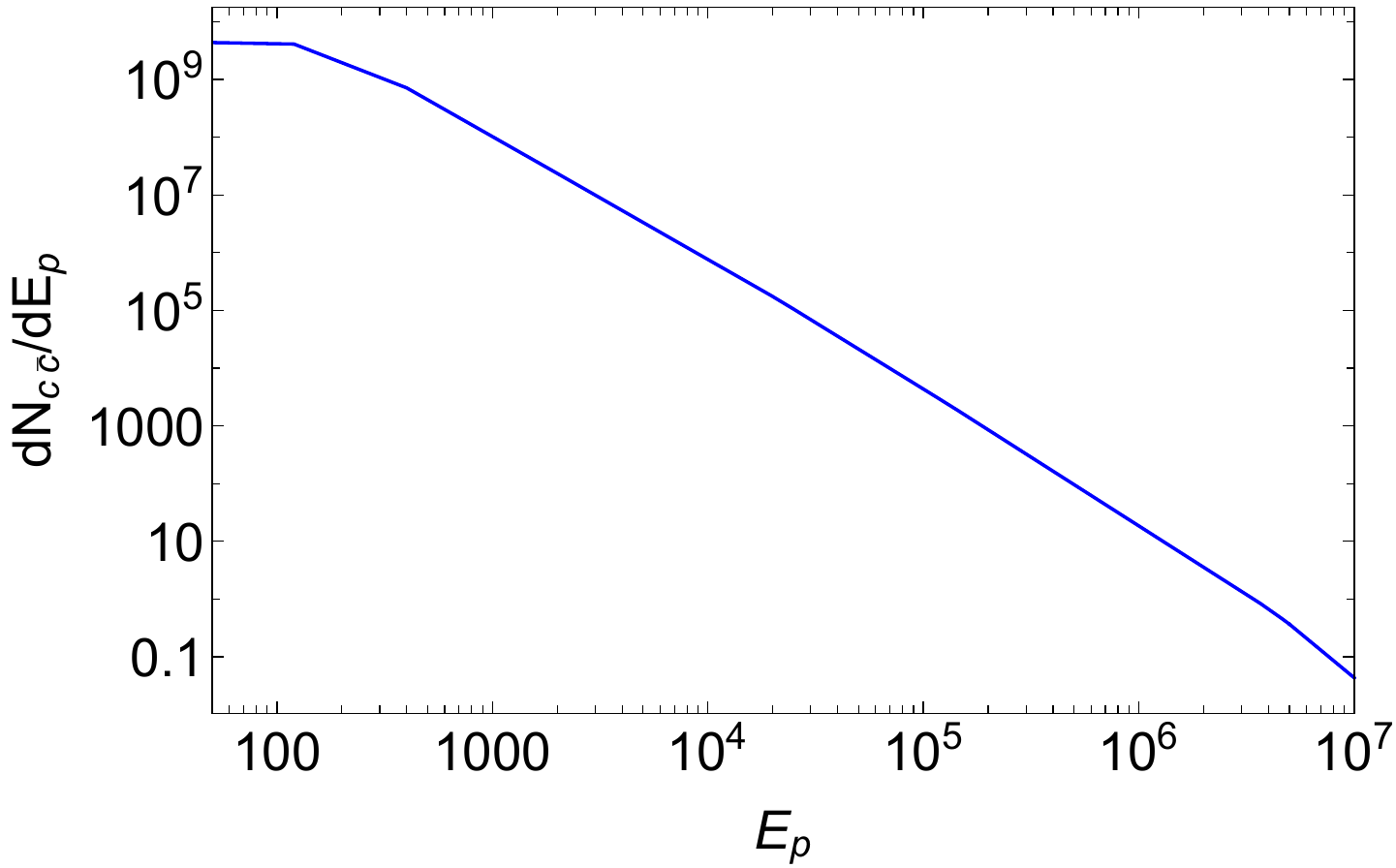}
    \caption{The integrand of Eq.~\eqref{eq:Ncc-KM3NeT}.}
    \label{fig:cc-flux-atmosphere}
\end{figure}

Now, let us discuss the sensitivity of KM3NeT to HNLs. Let us first compare the amount of HNL decay events at CHARM and KM3NeT for given value of the mixing angle at the lower bound of the sensitivity using simple analytic estimates. According to Eq.~\eqref{eq:nevents-main}, for the ratio of decay events at these experiments we have
\begin{multline}
    \frac{N_{\text{events,CHARM}}^{(\tau)}}{N_{\text{events,KM3NeT}}^{(\tau)}} \simeq \frac{N_{c\bar{c}}^{\text{CHARM}}\cdot \epsilon_{\text{geom}}^{\text{CHARM}}\cdot \epsilon^{\text{CHARM}}_{\text{decay}}}{N_{c\bar{c}}^{\text{KM3NeT}}}\times \frac{l_{\text{fid}}^{\text{CHARM}}}{l_{\text{fid}}^{\text{KM3NeT}}}\times\\  \times \frac{\gamma_{N}^{\text{KM3NeT}}}{\gamma_{N}^{\text{CHARM}}}\times \frac{\sum_{l = e,\mu}\Gamma(N_{\tau}\to ll)\epsilon_{\text{det},ll}}{\Gamma(N_{\tau}\to \mu\mu)}
    \label{eq:nevents-ratio-charm-km3net}
\end{multline}
Here, $N_{c\bar{c}}^{\text{CHARM}}\cdot \epsilon_{\text{geom}}^{\text{CHARM}}\cdot \epsilon_{\text{decay}}^{\text{CHARM}} \simeq 2\cdot 10^{13}$ (see Fig.~\ref{fig:epsilons} is the number of $c\bar{c}$ pairs detectable fraction of HNL decay events at CHARM. $N_{c\bar{c}}^{\text{KM3NeT}}$ is the amount of $c\bar{c}$ pairs produced in the upper hemisphere propagating to KM3NeT,
\begin{equation}
  N_{c\bar{c}}^{\text{KM3NeT}} \simeq 2\pi \times 1\text{ km}^{2}\times 5\text{ years}\times \int \frac{d\Phi}{d\Omega dtdS dE_{p}}\cdot \frac{\sigma_{pp\to c\bar{c}X}}{\sigma_{pp,\text{total}}}dE_{p} \simeq 10^{12},
  \label{eq:Ncc-KM3NeT}
\end{equation}
where $\sigma_{pp\to c\bar{c}X}(E_{p})$ is the energy-dependent charm production cross-section which we use from FONLL~\cite{Cacciari:2015fta} and from~\cite{Bezrukov:2009yw}, and $\sigma_{pp,\text{total}}$ is the total pp-cross-section, which we use from~\cite{Alekhin:2015byh}. The integrand in~\eqref{eq:Ncc-KM3NeT} is the product of two competing factors: $\frac{d\Phi}{d\Omega dtdS dE_{p}}$, which decreases with the proton's energy, and $\sigma_{pp\to c\bar{c}X}(E_{p})$, which increases, see Fig.~\ref{fig:cc-flux-atmosphere}.

We approximate the ratio of the mean HNL $\gamma$ factors by the ratio of the mean $\gamma$ factors of $D$ mesons:
\begin{equation}
\gamma_{N}^{\text{KM3NeT}}/\gamma_{D_{s}}^{\text{CHARM}} \simeq \gamma_{D_{s}}^{\text{KM3NeT}}/\gamma_{D_{s}}^{\text{CHARM}} \simeq 3,
\end{equation}
where we calculate $\gamma_{D_{s}}^{\text{KM3NeT}}$ using the $c\bar{c}$ distribution $\frac{d\Phi}{d\Omega dtdS dE_{p}}\cdot \sigma_{pp\to c\bar{c}X}$, assuming that $E_{D}\approx E_{p}/2$.

Using the fiducial lengths $l_{\text{fid}}^{\text{CHARM}} = \unit[35]{m}$ and $l_{\text{fid}}^{\text{KM3NeT}} \simeq \unit[1]{km}$, and taking into account that the last factor in Eq.~\eqref{eq:nevents-ratio-charm-km3net} is $\mathcal{O}(1)$ for $m_{N}\gg 2m_{\mu}$, we finally obtain
\begin{equation}
    \frac{N_{\text{events,CHARM}}^{(\tau)}}{N_{\text{events,KM3NeT}}^{(\tau)}} \simeq 2
\end{equation}
Therefore, naively, even in the most optimistic case (assuming unit efficiency) the number of events are just comparable. We need more accurate estimate taking into account non-isotropic distribution of the produced HNLs.

\subsection{Accurate estimate}
We compute the production of $D_{s}$ mesons (and hence $\tau$ leptons) using the approach from~\cite{Gondolo:1995fq}. The production was found to be maximal at $O(\unit[10]{km})$ height from the Earth's surface. The resulting spectrum $\frac{d\Phi_{D_{s}}}{dSdtdld\cos(\theta)dE_{D_{s}}}$ of $D_s$ mesons is in good agreement with Fig.~2 from~\cite{Arguelles:2019ziu}. The total number of $D_{s}$ mesons produced in the direction of KM3NeT during the operating time 5 years was found to be $N_{D_{s}}\simeq 5\cdot 10^{10}$.

Next, we use the approach from~\cite{Arguelles:2019ziu} in order to estimate the sensitivity of KM3NeT. The number of decay events is
        \begin{multline}
            N_{\text{events}} \approx S_{\text{Km3NeT}}\times T \times\int \frac{d\Phi_{D_{s}}}{dSdtdld\cos(\theta)dE_{D_{s}}}\cdot \text{Br}(D_{s}\to \tau\bar{\nu}_{\tau})\cdot\\ \cdot \text{Br}(\tau\to N_{\tau}X)\cdot  P_{\text{decay}}(l,E_{N})d\cos(\theta)dl dE_{N},
            \label{eq:1}
        \end{multline}
where $T = 5\text{ years}$ is the operating time, $S_{\text{KM3NeT}} = \unit[1]{km}^{2}$ is the transverse area of KM3NeT. The decay probability is
        \begin{equation}
            P_{\text{decay}} \approx e^{-(l+l_{1})/l_{\text{decay}}} - e^{-(l+l_{2})/l_{\text{decay}}},
        \end{equation}
where $l$ is the distance from the HNL production point in atmosphere, $l_{1} \approx \unit[3]{km}$ is the distance from the surface of Earth to the KM3NeT detector, while $l_{2} = l_{1}+\unit[1]{km}$ is the distance to the end of the KM3NeT. For simplicity, in $l_{\text{decay}}$ we set $E_{N}\approx E_{D_{s}}/2$. In order to show the maximal reach of KM3NeT, we optimistically assume unit efficiency of the dimuon event reconstruction, and require $N_{\text{events}}>3$ during the operating period.

The resulting sensitivity shown in Fig.~\ref{fig:results} is worse than predicted by the simple estimate by a factor of few. The reason is that at masses $m_{N}\lesssim \unit[500]{MeV}$ there is an additional suppression from $\text{Br}(N\to \mu\mu)$, while at higher masses the scaling~\eqref{eq:nevents-main} is not valid because the lower bound is close to the upper bound.

\bibliographystyle{JHEP}
\bibliography{bibliography}
\end{document}